\documentclass[twocolumn,showpacs,preprintnumbers,amsmath,amssymb]{revtex4}
\usepackage{graphicx}
\begin{document}
\renewcommand{\thefigure}{\arabic{figure}}
\title{Self-consistent theory of pair distribution functions and effective interactions in quantum Coulomb liquids}

\author{B. Davoudi$^{1,2}$, M. Polini$^{1}$, R. Asgari$^{1,2}$, and M. P. Tosi$^1$}
\affiliation{$^1$NEST-INFM and Classe di Scienze, Scuola Normale Superiore, I-56126 Pisa, Italy\\
$^2$Institute for Studies in Theoretical Physics and Mathematics, Tehran
19395-5531, Iran}

\begin{abstract}
We use a density-functional theoretical approach to set up a computationally simple self-consistent scheme to calculate the pair distribution functions and the effective interactions in quantum Coulomb liquids. 
We demonstrate the accuracy of the approach for different statistics and space dimensionalities by reporting results for a two-dimensional electron gas and for a three-dimensional boson plasma over physically relevant ranges of coupling strength, in comparison with Monte Carlo data.
\end{abstract}
\maketitle
Exchange and correlations in systems of interacting electrons have been a focus of interest in many-body physics for many decades. An important manifestation of these effects is the equilibrium distribution of electron pairs, which provides a picture of the short-range order in the system. Knowledge of this distribution for a homogeneous electron fluid as a function of its density determines its ground-state energy and is essential in the construction of energy functionals for the study of atoms, molecules and solids in applications of density functional theory (DFT) transcending a local-density approximation~\cite{Gunnarsson}. Recent developments in DFT have drawn attention to the utilization of inhomogeneous electron-pair densities for such studies~\cite{Gonis}. There also is an interest for fluids of charged bosons in quantum statistical mechanics, {\it e.g.} in regard to condensates of point-like Cooper pairs as models for superfluid states~\cite{Mott}.

We present in this Letter a novel theory of the pair distribution function $g(r)$ in homogeneous fluids of charged, point-like fermions or bosons. We start from some basic results of DFT~\cite{Dreizler} and tackle the evaluation of $g(r)$ through the solution of two-particle scattering problems which are governed by effective scattering potentials embodying the many-body effects. This viewpoint, which was first taken by Overhauser~\cite{Overhauser} and further developed in subsequent studies~\cite{Gp,Davoudi}, is here implemented through a self-consistent inclusion of exchange and correlations leading back, when a non-self-consistent linear-response approximation is made, to the spin-dependent effective interactions derived in the early work of Kukkonen and Overhauser~\cite{Kukkonen}. Calculations on fluids of electrons  in dimensionality $D=2$ and of charged bosons in $D=3$ are shown to yield results in excellent agreement with available Quantum Monte Carlo data over a physically significant range of coupling strength. As previously noted, the method can be extended to pair distributions in inhomogeneous Coulomb fluids~\cite{Davoudi}.

We consider a quantum fluid of point-like particles having charge $e$ and mass $m$ at zero temperature, neutralized by a uniform charged background.
The fluid consists of different components at average density $n_{\alpha}$ (two spin components for spin-$1/2$ fermions, for instance).  As already stated, our aim is to use DFT for building a self-consistent theory that gives the pair distribution functions as output. From the Hohenberg-Kohn theorem~\cite{Dreizler}, the ground-state energy of the fluid in the presence of a set of external potentials $V^{\rm ext}_{\alpha}({\bf r})$ can be written as
\begin{eqnarray}\label{gsf}
E_{\rm gs}[\{n_{\sigma}({\bf r})\}]=T_s[\{n_{\sigma}({\bf r})\}]+E_{\scriptscriptstyle \rm H}[\{n_{\sigma}({\bf r})\}]&&\nonumber\\
+\sum_{\alpha}\int\,d^D{\bf r}\,\,V^{\rm ext}_{\alpha}({\bf r})\Delta n_{\alpha}({\bf r})+E_{\scriptscriptstyle \rm Q}[\{n_{\sigma}({\bf r})\}]&&
\end{eqnarray}
where $\{n_{\sigma}({\bf r})\}$ is the set of densities of all components, $\Delta n_{\alpha}({\bf r})\equiv n_{\alpha}({\bf r})-n_{\alpha}$, and  $T_s$ is the  ideal kinetic energy functional. The Hartree term $E_{\scriptscriptstyle \rm H}$ is given by
\begin{equation}\label{hartree}
E_{\rm H}=\frac{1}{2}\sum_{\alpha,\beta}\int d^D{\bf r}\int d^D{\bf r}'\,v(|{\bf r}-{\bf r}'|)\,\Delta n_{\alpha}({\bf r})\Delta n_{\beta}({\bf r}')
\end{equation}
where $v(|{\bf r}-{\bf r}'|)=e^2/|{\bf r}-{\bf r}'|$. The last term in Eq.~(\ref{gsf}) is the exchange-correlation energy functional, containing the quantum many-body (QMB) effects. In Eqs.~(\ref{gsf}) and (\ref{hartree}) the presence of a neutralizing background has been taken into account. 

The quantity $n_{\alpha}\,[g_{\alpha \gamma}(r)-1]$, where $g_{\alpha \gamma}(r)$ is the component-resolved pair distribution function, can be viewed as the distortion that a particle of the fluid (of $\gamma$-type at position ${\bf r}=0$) produces in the density profiles~\cite{Percus}. Here, $g_{\alpha \gamma}(r)$ is defined through the average number of $\alpha$-type particles inside a spherical shell of radius $r$ and thickness $dr$ centered on a $\gamma$-type particle located at the origin, which is given by $n_{\alpha}g_{\alpha\gamma}(r)\Omega_Dr^{D-1}dr$ with $\Omega_2=2\pi$ and $\Omega_3=4\pi$. The appropriate ground-state energy functional is obtained from Eq.~(\ref{gsf}) with the formal replacements $V^{\rm ext}_{\alpha}({\bf r}) \rightarrow v(r)$ and $\Delta n_{\alpha}({\bf r}) \rightarrow n_{\alpha}\,[g_{\alpha \gamma}(r)-1]$. Finally, the QMB energy functional can be written {\it via} an adiabatic connection formula,
\begin{eqnarray}\label{acf}
E^{(\gamma)}_{\scriptscriptstyle \rm Q}[\{n_{\sigma}(r)\}]\!\!\!&\!\!=\!\!&\!\!\!
\frac{1}{2}\sum_{\alpha,\beta}\frac{1}{e^2}\int_{0}^{e^2} d \lambda
\int d^D{\bf r}\int d^D{\bf r}'\,v(|{\bf r}-{\bf r}'|)\nonumber \\ 
&\times &n_{\alpha}(r)n_{\beta}(r')  \left[g^{\lambda}_{\alpha\beta\gamma}({\bf r}, {\bf r}')-1\right]~,
\end{eqnarray}
where $n_{\alpha}(r)=n_{\alpha} g_{\alpha\gamma}(r)$ and $g^{\lambda}_{\alpha\beta\gamma}({\bf r}, {\bf r}')$ measures the probability of finding two particles with indices $\alpha$ and $\beta$ at ${\bf r}$ and ${\bf r}'$ when a $\gamma$-type particle is at the origin, the interaction potential being  $v_{\lambda}(|{\bf r}-{\bf r}'|)=\lambda\,e^2/|{\bf r}-{\bf r}'|$. Of course,  $g^{\lambda}_{\alpha\beta\gamma}({\bf r}, {\bf r}')$ depends functionally on $\{n_{\sigma}({\bf r})\}$.

The Kohn-Sham mapping~\cite{Dreizler} ensures that $n_{\alpha}(r)$ can be built from Kohn-Sham scattering orbitals $\Phi^{\alpha \gamma}_{\bf k}({\bf r})$ which satisfy the following set of Schr\"{o}dinger equations:
\begin{equation}\label{sks}
\left[-\frac{\hbar^2}{2 \mu}\nabla^2_{\bf r}+V^{\alpha \gamma}_{\scriptscriptstyle \rm KS}(r)\right]\Phi^{\alpha \gamma}_{\bf k}({\bf r})=\varepsilon_{{\bf k}}\,\Phi^{\alpha \gamma}_{\bf k}({\bf r})~.
\end{equation}
Here, ${\bf r}$ is the relative distance of two particles, $\mu=m/2$ is the reduced mass, and 
$\varepsilon_{\bf k}= \hbar^2 k^2/2 \mu$ with $\hbar {\bf k}$ the relative momentum. The scattering potential $V^{\alpha \gamma}_{\scriptscriptstyle \rm KS}(r)$ in Eq.~(\ref{sks}) can be obtained from the first functional derivative of $E_{\rm gs}-T_s$ with respect to $n_{\alpha}(r)$,
\begin{equation}\label{ks}
V^{\alpha \gamma}_{\scriptscriptstyle \rm KS}(r)=v(r)+\sum_{\beta}\int d^D{\bf r}' v(|{\bf r}-{\bf r}'|)\Delta n_{\beta}({\bf r}')+\frac{\delta E^{(\gamma)}_{\scriptscriptstyle \rm Q}}{\delta n_{\alpha}(r)}.
\end{equation}
Finally, the pair distribution functions can be obtained from the Kohn-Sham scattering states as
\begin{equation}\label{gofr}
g_{\alpha\gamma}(r)=\sum_{{\bf k},\, {\rm occ.}}\Gamma^{\alpha\gamma}_{\bf k}\,|\Phi^{\alpha \gamma}_{\bf k}({\bf r})|^2
\end{equation}
where the sum runs over all occupied states labelled by ${\bf k}$. The occupation factors $\Gamma^{\alpha\gamma}_{\bf k}$ depend on the statistics of the fluid (see the discussion below). Notice that Eq.~(\ref{gofr}) guarantees positive definiteness of $g_{\alpha\gamma}(r)$.

In the above formal development, the functional dependence of the QMB energy on density is not known and we have to resort at this point to some approximations. Their goodness can only be gauged {\it a posteriori}.  Firstly, the function $g^{\lambda}_{\alpha\beta\gamma}({\bf r}, {\bf r}')$ in Eq.~(\ref{acf}) involves three-body correlations and would lead us into a hierarchy of higher-order correlation functions. The simplest way of truncating this hierarchy is to replace $g^{\lambda}_{\alpha\beta\gamma}({\bf r}, {\bf r}')$ by $g^{\lambda}_{\alpha\beta}(|{\bf r}-{\bf r}'|)$, in analogy with what has been done in treating the equation of motion for the Wigner distribution function in the presence of external potentials~\cite{Singwi}. Secondly, we expand the QMB energy in a functional Taylor series in powers of $\Delta n_{\alpha}(r)$ up to second order terms. With the definition
\begin{equation}\label{fxc}
{\rm f}^{\alpha\beta}(|{\bf r}-{\bf r}'|)\equiv \left.\frac{\delta^{2} E_{\scriptscriptstyle \rm Q}[\{n_{\sigma}(r)\}]}{\delta n_{\alpha}(r)\delta n_{\beta}(r')}\right|_{\{\Delta n_{\sigma}(r)\}=0}
\end{equation}
we find in Fourier transform
\begin{equation}\label{pot}
V^{\alpha \gamma}_{\scriptscriptstyle \rm KS}(q)=v(q)+\sum_{\beta}v(q)[1-G_{\alpha\beta}(q)]\,\Delta n_{\beta}(q)~.
\end{equation}
Here, $G_{\alpha\beta}(q)\equiv -f^{\alpha \beta}(q)/v(q)$ are the so-called local field factors, defined in term of the Fourier transform $f^{\alpha \beta}(q)$ of the QMB kernels in Eq.~(\ref{fxc}), and $v(q)$ is the Fourier transform of the Coulomb potential ({\it i.e.} $v(q)= 4 \pi e^2/q^2$ in $D=3$, $v(q)= 2 \pi e^2/q$ in $D=2$). It may be worth exploring in the future  alternative approximations to a truncated expansion of the QMB energy.

The approximations that have led us to Eq.~(\ref{pot}) can be justified in a weak-coupling regime and indeed Eq.~(\ref{pot}) yields back the effective electron-electron interactions of Kukkonen and Overhauser~\cite{Kukkonen} when $\Delta n_{\beta}(q)$ is related to the scattering potential $V^{\alpha \gamma}_{\scriptscriptstyle \rm KS}(q)$ by linear response theory. We propose instead to carry out a self-consistent calculation of the pair distribution functions and of the effective interactions through the solution of the set of equations (\ref{sks}), (\ref{gofr}) and (\ref{pot}). For this purpose the quantities $\Delta n_{\beta}(q)$ in Eq.~(\ref{pot}) should be written in the form
\begin{equation}\label{pft}
\Delta n_{\beta}(q)= (n_\beta/n_\gamma)^{1/2}\,[S_{\beta\gamma}(q)-\delta_{\beta\gamma}]
\end{equation}
where $S_{\beta\gamma}(q)$ are the partial structure factors, related to the pair functions by
\begin{equation}
S_{\beta\gamma}(q)=\delta_{\beta\gamma}+\sqrt{n_{\beta}n_{\gamma}}\,\int d^{D} {\bf r}\,[g_{\beta\gamma}(r)-1]\exp{(-i {\bf q} \cdot {\bf r})}~.
\end{equation}
The uselfulness of such a self-consistent approach will be explored in the numerical calculations reported further below. Here we remark that setting $G_{\alpha\beta}(q)=0$ in Eq.~(\ref{pot}) gives back the self-consistent Hartree approximation (HA), that we have found to yield quite satisfactory results for $g(r)$ in the $3D$ electron gas for values of the coupling strength up to at least $r_s =10$~\cite{Davoudi}. 

Let us examine the application of the above self-consistent scheme (SCS) to a $2D$ paramagnetic electron gas (EG), where correlations are stronger than in $3D$. The Greek indices become spin indices taking the values $\sigma=\pm 1$. For the details of the summation procedure in Eq.~(\ref{gofr}), the reader is referred to Refs.~\cite{Gp} and \cite{Davoudi}. The necessary input are the local-field factors $G_{\sigma\sigma'}(q)$, for which we use Quantum Monte Carlo (QMC) data~\cite{Saverio1} as described by interpolation formulae in Ref.~\cite{Giuliani}. These refer to the charge-charge and spin-spin field factors, $G_{+}(q)=[G_{\uparrow\uparrow}(q)+G_{\uparrow \downarrow}(q)]/2$ and $G_{-}(q)=[G_{\uparrow\uparrow}(q)-G_{\uparrow \downarrow}(q)]/2$, which in the long-wavelength limit satisfy the compressibility and susceptibility sum rules, $G_{+}(q)\rightarrow (\kappa^{-1}_0-\kappa^{-1})/[n^2\,v(q)]$ and $G_-(q)\rightarrow \mu^2_B(\chi^{-1}_{\rm P}-\chi^{-1}_s)/v(q)$. Here, $\kappa_0$ and $\chi_{\rm P}$ are the compressibility and the Pauli susceptibility of the ideal Fermi gas, $\kappa$ and $\chi_s$ are the corresponding quantities for the EG, and $\mu_B$ is the Bohr magneton. We recall that these expressions, though strictly valid only in the thermodynamic limit, give in practice a good account of $G_{\pm}(q)$ over a range of $q$ extending almost up to $2 k_F$~\cite{Saverio1,Giuliani}.

\begin{figure*}
\includegraphics[scale=0.7]{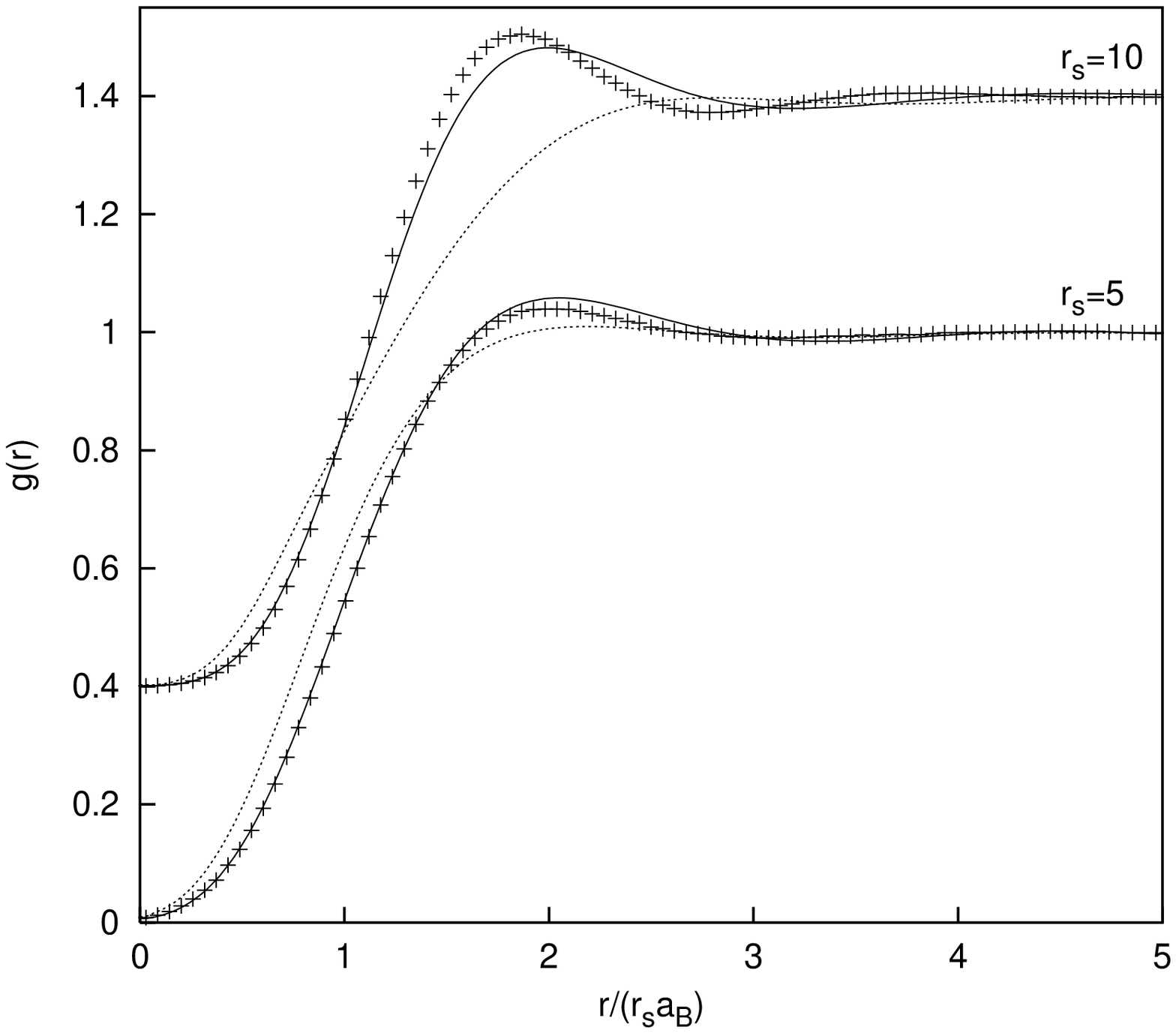}
\caption{The pair distribution function $g(r)$ in a $2D$ EG at $r_s=5$ and $10$, as a function of $r/(r_s a_B)$. The results of the SCS (full line) and of the HA (dotted line) are compared with QMC data (crosses). 
The curves at $r_s=10$ have been shifted upwards by 0.4.}
\label{Fig1}
\end{figure*}

Figure \ref{Fig1} reports our SCS results for the spin-averaged pair distribution function $g(r)=[g_{\uparrow\uparrow}(r)+g_{\uparrow\downarrow}(r)]/2$ in the $2D$ EG at $r_s=5$ and $10$, where $r_s=(\pi n a^2_B)^{-1/2}$ is the usual coupling-strength parameter with $a_B$ the Bohr radius. The results of a QMC study by S. Moroni (private communication) and of our previous HA approach are also shown in Figure \ref{Fig1}. It is seen that inclusion of exchange and correlation in the present SCS reproduces the formation of a first-neighbor shell with increasing coupling strength, which is missed in the HA~\cite{Davoudi}, and yields satisfactory quantitative agreement with the QMC data.

\begin{figure*}
\includegraphics[scale=0.7]{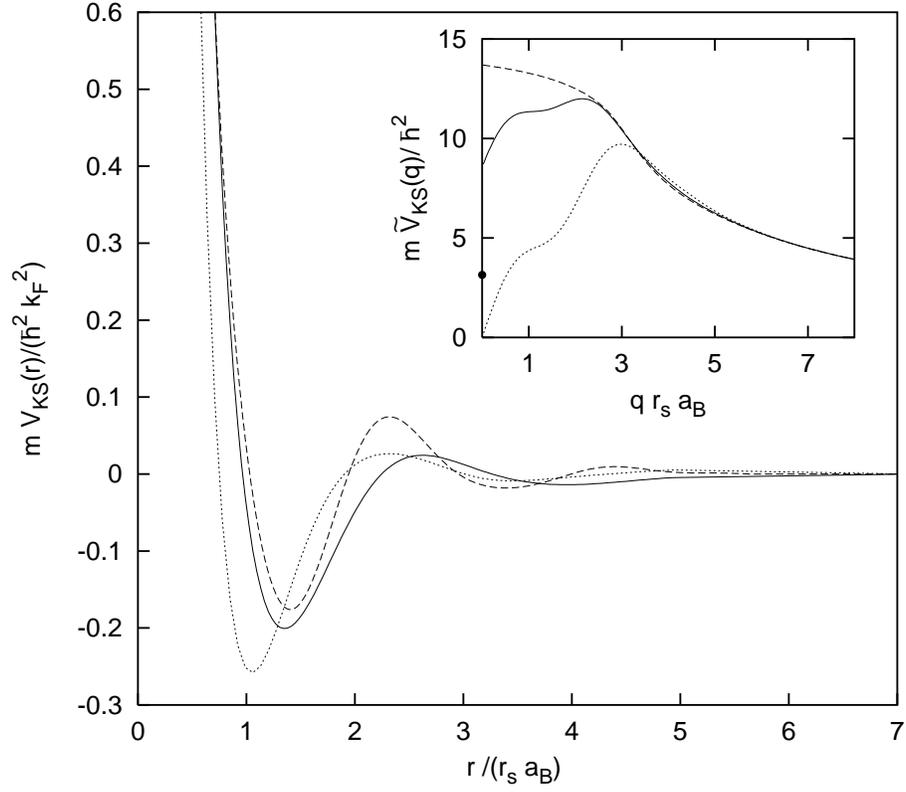}
\caption{The spin-symmetric effective electron-electron interaction $V_{\scriptstyle \rm KS}(r)$ in a $2D$ EG at $r_s=5$, as a function of $r/(r_s a_B)$. The result of the SCS (full line) is compared with the HA (dotted line) and with the Kukkonen-Overhauser result (dashed line). 
The inset shows the Fourier transform of $V_{\scriptstyle \rm KS}(r)$
as a function of $q r_s a_B$. The black dot gives the long-wavelength limit of the Thomas-Fermi theory for the electron-test charge interaction.}
\label{Fig2}
\end{figure*}

Figure \ref{Fig2} reports the SCS results for the spin-symmetric component of the effective electron-electron interaction, $V_{\scriptstyle \rm KS}(r)=[V^{\uparrow \uparrow}_{\scriptstyle \rm KS}(r)+V^{\uparrow \downarrow}_{\scriptstyle \rm KS}(r)]/2$, for the $2D$ EG at $r_s=5$, as well as its Fourier transform ${\widetilde V}_{\scriptstyle \rm KS}(q)$ (shown in the inset). We find very significant changes from the HA and indeed the original treatment of exchange and correlations in a linear-response approximation by Kukkonen and Overhauser~\cite{Kukkonen} are not very far from our SCS results.

We should remark at this point that the present SCS	is not as accurate in reproducing the spin-resolved pair distribution functions of the $2D$ EG and hence the spin-spin effective interaction. Numerical evidence that an empirical inclusion of higher-order terms in the scattering potentials can yield full agreement with the QMC data, as well as extensions to larger values of $r_s$, will be reported in future work.

Here we discuss instead how our approach can be extended into a fully self-consistent scheme (FSCS), in which the local field factors are self-consistently determined over the relevant $q$-range during the calculation rather than taken as input from QMC. We need for this purpose a closure relation between $G_{\alpha\beta}(q)$ and $S_{\alpha\beta}(q)$, and the crucial point is that it should self-consistently satisfy the thermodynamic (compressibility and susceptibility) sum rules. 
Such a closure can be obtained by using Eq.~(\ref{acf}) in Eq.~(\ref{fxc}), with the aforementioned approximation $g^{\lambda}_{\alpha\beta\gamma}({\bf r}, {\bf r}') \simeq g^{\lambda}_{\alpha\beta}(|{\bf r}-{\bf r}'|)$. The details of this calculation will be reported elsewhere. The final result is
\begin{equation}\label{fattorilocali}
G_{\alpha\beta}(q)=D_{\alpha\beta}\,{\mathcal G}_{\alpha\beta}(q)~.
\end{equation}
Here, the differential operator $D_{\alpha\beta}$ is defined by
\begin{equation}
D_{\alpha\beta}\equiv 1+n_{\alpha}\frac{\partial}{\partial n_{\alpha}}+n_{\beta}\frac{\partial}{\partial n_{\beta}}+\frac{1}{2}n_{\alpha}n_{\beta}\,\frac{\partial^2}{\partial n_{\alpha}\partial n_{\beta}}~,
\end{equation}
while ${\mathcal G}_{\alpha\beta}(q)$ is given by 
\begin{eqnarray}\label{gstorto}
\!\!{\mathcal G}_{\alpha\beta}(q)\equiv &&-\frac{1}{\sqrt{n_\alpha n_\beta}}\,\frac{1}{e^2}\,\int_{0}^{e^2} d\lambda\, \int \frac{d^{D} {\bf q}'}{(2 \pi)^D}\,\frac{v(q')}{v(q)}\nonumber\\
&&\times\left[S^{\lambda}_{\alpha\beta}(|{\bf q}+{\bf q}'|)-\delta_{\alpha\beta}\right]~,
\end{eqnarray}
with $S^{\lambda}_{\alpha\beta}(q)$ being the partial structure factor at coupling constant $\lambda$. Although these expressions are strictly correct only in the long-wavelength limit, they yield a good account of QMC data on the local field factors over the relevant range of $q$~\cite{Saverio1,Giuliani}, as already noted.

It can be seen from Eqs.~(\ref{fattorilocali})-(\ref{gstorto}) that, in order to satisfy the susceptibility sum rule in the paramagnetic EG, one needs to move into a partially spin-polarized state and indeed evaluate the full range of spin polarization up to the ferromagnetic state. For a simple illustration of our method we have therefore considered a $3D$ fluid of spinless charged bosons (CBF), where the proposed FSCS simplifies drastically. At zero temperature all bosons in the reference Kohn-Sham ideal gas are in the $k=0$ state, so that $g(r) \propto |\Phi_{{\bf k}={\bf 0}}({\bf r})|^2$. The scattering orbital $\Phi_{{\bf k}={\bf 0}}({\bf r})$ is a spherically symmetric function and the Kohn-Sham Schr\"odinger equation becomes  equivalent to the Euler-Lagrange equation for $g(r)$ as obtained from the variational principle using the  von Weizs\"acker functional for $T_s$~\cite{von}, $T_s[g(r)]=(\hbar^2 n/8 \mu)\,\int d^D{\bf r}\,|\nabla g(r)|^2/g(r)$. Furthermore, the local field factor $G(q)$ from Eqs.~(\ref{fattorilocali})-(\ref{gstorto}) satisfies the compressibility sum rule, {\it i.e.} $\lim_{q \rightarrow 0} G(q)=-[n^2 \kappa\, v(q)]^{-1}$ with $\kappa$ being the compressibility of the interacting Bose liquid. 

\begin{figure*}
\includegraphics[scale=0.7]{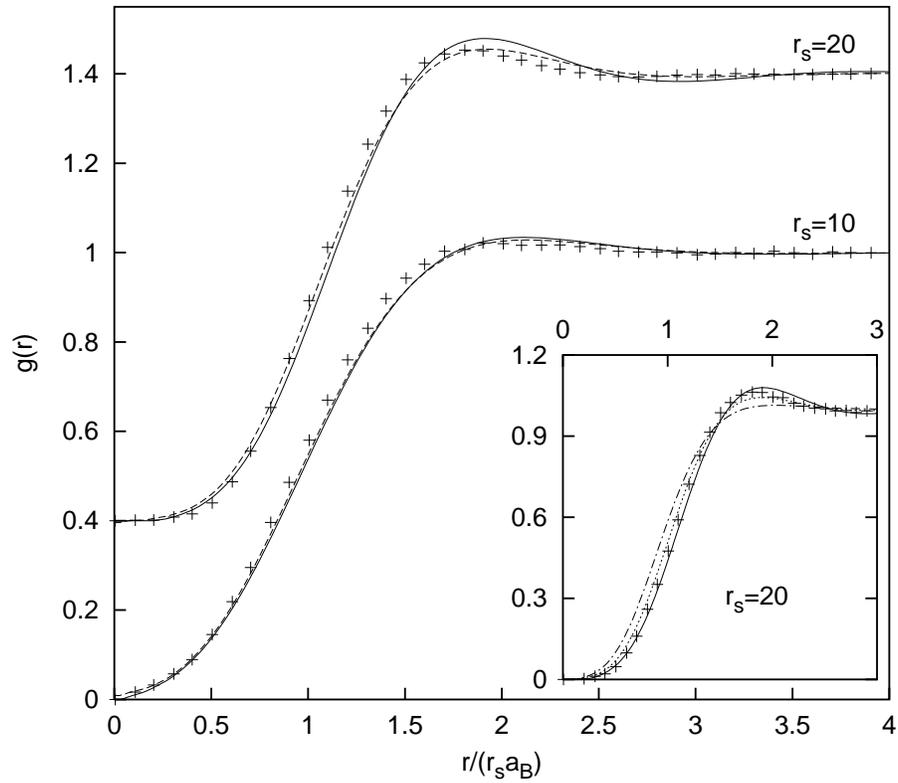}
\caption{The pair distribution function $g(r)$ in a $3D$ CBF at $r_s=10$ and $20$, as a function of $r/(r_s a_B)$. The results of the FSCS (full line) and of the SCS (dashed line) are compared with QMC data (crosses). The curves at $r_s=20$ have been shifted upwards by 0.4. In the inset, the HA result (dash-dotted line) is compared with the FSCS (full line), the QMC data (crosses), and the HNC of Apaja {\it et al.} (dotted line).}
\label{Fig3}
\end{figure*}

\begin{figure*}
\includegraphics[scale=0.7]{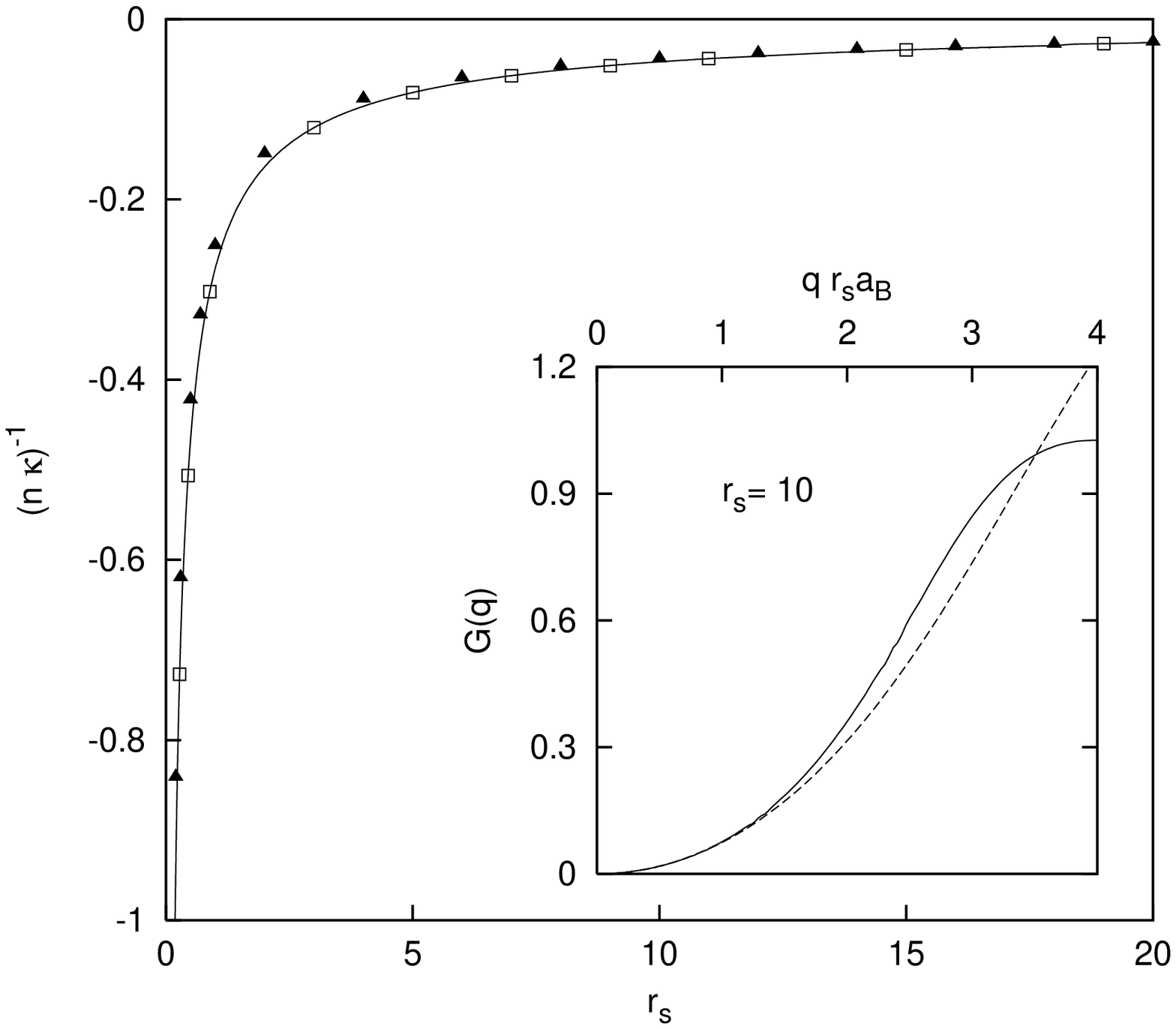}
\caption{$(n\,\kappa)^{-1}$ in Ryd as a function of $r_s$ for a $3D$ CBF. The results of the FSCS from the ground-state energy (full line) and from the self-consistent $G(q)$ (empty boxes) are compared with QMC data (triangles). 
The inset shows $G(q)$ as a function of $q\, r_s a_B$ at $r_s=10$: 
the FSCS (full line) is compared with QMC data (dashed line).}
\label{Fig4}
\end{figure*}

We have solved the FSCS based on Eqs.~(\ref{sks}), (\ref{gofr}), (\ref{pot}) and (\ref{fattorilocali})-(\ref{gstorto}) for a 
$3D$ CBF at coupling strength $r_s\equiv(4 \pi n a^3_B/3)^{-1/3}$ up to $20$. The main results are shown in Figures \ref{Fig3} and \ref{Fig4}. In Figure \ref{Fig3} we compare the FSCS $g(r)$ with QMC data by Moroni {\it et al}.~\cite{Moroni} and with SCS results where the QMC data for the local field factor~\cite{Moroni} have been used as input. Our results for $g(r)$ are in excellent agreement with QMC and with each other, the implication being that our self-consistent determination of $G(q)$ from the compressibility sum rule also accounts with sufficient accuracy for this function over the relevant range of $q$ (see also the inset in Figure \ref{Fig4}, where the FSCS $G(q)$ is compared with the QMC data). In the inset in Figure \ref{Fig3} we include, with the FSCS results and the QMC data for $g(r)$ at $r_s=20$, also the results obtained in the HA and those of Apaja {\it et al}.~\cite{Mott} based on a hypernetted chain (HNC) approximation. Again, the HA is not able to reproduce quantitatively the formation of a first-neighbor peak with increasing coupling. Finally, in the main body of Figure \ref{Fig4} we illustrate the self-consistency that we have obtained in the values of the compressibility $\kappa$ from the ground-state energy and from $G(q)$, as well as their excellent agreement with QMC data by Moroni {\it et al.}~\cite{Moroni}. 

In summary, we have proposed a self-consistent approach by which the pair correlations (and related quantities such as the internal energy) of a quantum Coulomb fluid can be determined, ultimately using the partial mean densities of its components as the only input. We have examined its usefulness in different statistics (fermions and bosons) and space dimensionalities ($D=2$ and $3$) in comparison with available QMC data, extending up to relatively large values of the Coulomb coupling strength. We have also highlighted the role of exchange and correlations in determining the emergence of liquid-like structure with increasing coupling strength through the formation of a first-neighbor shell and further oscillations in the pair distribution function. 

This work was partially supported by MIUR through the PRIN2001 program. 
We are grateful to Prof. G. Vignale for stimulating discussions and to Dr. S. Moroni for giving us access to his QMC results for the $2D$ EG prior to publication.


\begin{references}
\bibitem{Gunnarsson}
O. Gunnarsson, M. Jonson, and B. I. Lundqvist, Phys. Rev. B {\bf 20}, 3136 (1979); J. P. Perdew and A. Zunger, {\it ibid.} {\bf 23}, 5048 (1981); E. Chac\`on and P. Tarazona, {\it ibid.} {\bf 37}, 4013 (1988); J. P. Perdew, K. Burke, and Y. Wang, {\it ibid.} {\bf 54}, 16533 (1996); J. P. Perdew, K. Burke, and M. Ernzerhof, Phys. Rev. Lett. {\bf 77}, 3865 (1996) and {\bf 78}, 1396 (1997).
\bibitem{Gonis}
A. Gonis, T. C. Schulthess, J. van Ek, and P. E. A. Turchi, Phys. Rev. Lett. 77, 2981 	(1996);
M. Levy and P. Ziesche, J. Chem. Phys. {\bf 115}, 9110 (2001). 
\bibitem{Mott}
A. S. Alexandrov and N. F. Mott, Supercond. Sci. Technol. {\bf 6}, 215 (1993); 
V. Apaja, J. Halinen, V. Halonen, E. Krotscheck, and M. Saarela, Phys. Rev. B {\bf 55}, 12925 (1997).
\bibitem{Dreizler}
R. M. Dreizler and E. K. U. Gross, {\it Density Functional Theory, An Approach to the 
Quantum Many-Body Problem} (Springer, Berlin, 1990).
\bibitem{Overhauser}
A. W. Overhauser, Can. J. Phys. {\bf 73}, 683 (1995).
\bibitem{Gp}
P. Gori-Giorgi and J. P. Perdew, Phys. Rev. B {\bf 64}, 155102 (2001).
\bibitem{Davoudi}
B. Davoudi, M. Polini, R. Asgari, and M. P. Tosi, to appear in Phys. Rev. B and cond-mat/0205339.
\bibitem{Kukkonen}
C. A. Kukkonen and A. W. Overhauser, Phys. Rev. B {\bf 20}, 550 (1979); G. Vignale and K. S. Singwi, Phys. Rev. B {\bf 32}, 2156 (1985); S. Yarlagadda and G. F. Giuliani, Phys. Rev. B {\bf 49}, 7887 (1994). 
\bibitem{Percus}
The classical analog of this prescription was first considered by J. K. Percus, in {\it The Equilibrium Theory of Classical Fluids}, edited by H. L. Frisch and J. L. Lebowitz (Benjamin, New York, 1964), p. II-83.
\bibitem{Singwi}
K. S. Singwi, 
M. P. Tosi, R. H. Land, and A. Sj\"olander, Phys. Rev. {\bf 176}, 589 (1968); R. Lobo, K. S. Singwi,
and M. P. Tosi, {\it ibid.} {\bf 186}, 470 (1969); P. Vashishta and K. S. Singwi, Phys. Rev. B {\bf 6}, 875 (1972).
\bibitem{Saverio1}
S. Moroni, D. M. Ceperley, and G. Senatore, Phys. Rev. Lett. {\bf 69}, 1837 (1992) and {\bf 75}, 689 (1995). 
\bibitem{Giuliani}
B. Davoudi, M. Polini, G. F. Giuliani, and M. P. Tosi, Phys. Rev. B {\bf 64}, 153101 and 233110 (2001).
\bibitem{von}
C. F. von Weizs\"acker, Z. Phys. {\bf 96}, 431 (1935); 
C. Herring, Phys. Rev. A {\bf 34}, 2614 (1986). 
\bibitem{Moroni}
S. Moroni, S. Conti, and M. P. Tosi, Phys. Rev. B. {\bf 53}, 9688 (1996).
\end{references}
\end{document}